\documentclass[twocolumn,showpacs,preprintnumbers,amsmath,amssymb,aps,prl]{revtex4}

\usepackage{epsfig}
\usepackage{graphicx}
\usepackage{dcolumn}
\usepackage{bm}

\begin{document}

\title{Evidence of the influence of magnetism on pseudogap states in the high
resolution spectra of EuFe$_2$As$_2$}

\author{Ganesh Adhikary, Nishaina Sahadev, Deep Narayan Biswas, R. Bindu, Neeraj Kumar,
A. Thamizhavel, S. K. Dhar and Kalobaran Maiti}
\altaffiliation{Corresponding author: kbmaiti@tifr.res.in}

\affiliation{Department of Condensed Matter Physics and Materials
Science, Tata Institute of Fundamental Research, Homi Bhabha Road,
Colaba, Mumbai - 400 005, INDIA.}

\begin{abstract}
Employing {\it state of the art} high resolution photoemission
spectroscopy, we studied the temperature evolution of the electronic
structure of EuFe$_2$As$_2$, an unique pnictide, where
antiferromagnetism of Eu layer survives within the superconducting
phase due to `FeAs' layers achieved via substitution and/or
pressure. High energy and angle resolution helped to reveal
pseudogap-quasiparticle features having primarily As 4$p$ character
and spin density wave transition induced band folding in the
electronic structure. A weakly dispersing feature of dominant As
4$p$ character is discovered around 80 meV that becomes weaker in
intensity below 20 K manifesting influence of antiferromagnetic
order on conduction electrons. These results provide an evidence of
a link between the pseudogap states and magnetism that could be
revealed employing high resolutions.
\end{abstract}

\date{\today}


\maketitle

Interplay of magnetism and superconductivity is a long standing
issue in material research and believed to play the key role in
deriving the unconventional superconductivity
\cite{spin-fluctuation}. Discovery of superconductivity in Fe-based
compounds \cite{Kamihara1,Kamihara2} via suppression of spin density
wave (SDW) state, further reinforces this connection and are
believed to be good candidates for such studies. These
superconductors are usually classified into the following categories
based on their composition and structure, namely `1111', `122',
`111' and `11' families \cite{Iron-Pnictide}. `FeAs' layers play the
key role in deriving the materials properties of these systems.
Among them, `122' class of materials, $A$Fe$_{2}$As$_{2}$ ($A$ = Ba,
Sr, Ca and Eu) have drawn significant attention as high quality
single crystals can be grown easily in the whole composition range
and one can easily tailor the properties of the materials via
insertion of varied elements in the layers intermediate to the
`FeAs' layers.

EuFe$_{2}$As$_{2}$ is a special compound in this class as the spacer
atom, Eu has large moment leading to two magnetic transitions, (i)
SDW-type antiferromagnetic transition around 190 K due to Fe-moments
and (ii) antiferromagnetic (AFM) ordering of the Eu-moments at 20 K.
The versatility of EuFe$_{2}$As$_{2}$ is that superconductivity (SC)
can be achieved by substituting foreign elements in any of the
sites. For example, substitution of K \cite{K-doped1,K-doped2} and
Na \cite{Na-doped} at Eu sites leads to superconductivity with
transition temperature, $T_C$ as high as 30 K and 34.7 K,
respectively. Fe site substitution by Co results in
superconductivity \cite{Co-doped1,Co-doped2}. Ni substitution at Fe
sites is multifaceted; while Ni-doped CaFe$_2$As$_2$ is
superconducting \cite{Ni-doped-CaFe2As2}, Eu moments align
ferromagnetically in Ni-doped EuFe$_2$As$_2$ at low temperatures and
no superconductivity was observed \cite{Ni-doped-EuFe2As2}. The
phase diagram of P-doping at As sites is also interesting
\cite{P-doped1,P-doped2}; a narrow low-doped regime exhibit
coexistence of superconductivity and AFM order, while a somewhat
higher doping leads to ferromagnetic order of Eu-moments and
superconductivity disappears. Application of pressure leads to
superconductivity even in the parent compound \cite{pressure}.
Evidently, EuFe$_2$As$_2$ provides an ideal playground to achieve
insight on the interplay of superconductivity and magnetism. Here,
we studied the evolution of the electronic structure as a function
of temperature and photon energy employing high resolution
photoemission spectroscopy. Experimental results provide signature
of pseudogap, features related to magnetic order and evidence of a
link between them.

High quality single crystalline, EuFe$_{2}$As$_{2}$ was grown using
Sn-flux\cite{growth}. The photoemission measurements were carried
out on cleaved sample surface using monochromatic photon sources and
Gammadata Scienta, R4000 WAL analyzer at a base pressure better than
3$\times$10$^{-11}$ torr. The temperature variation down to 10 K on
the sample was achieved using an open cycle liquid helium cryostat,
LT-3M from Advanced Research Systems, USA. The energy resolution was
fixed to 2 meV for angle integrated photoemission (AIPES) and 10 meV
for angle-resolved photoemission spectroscopy (ARPES) with angle
resolution set to $0.3^{\circ}$.

\begin{figure}
 \vspace{-4ex}
\includegraphics[scale=.45]{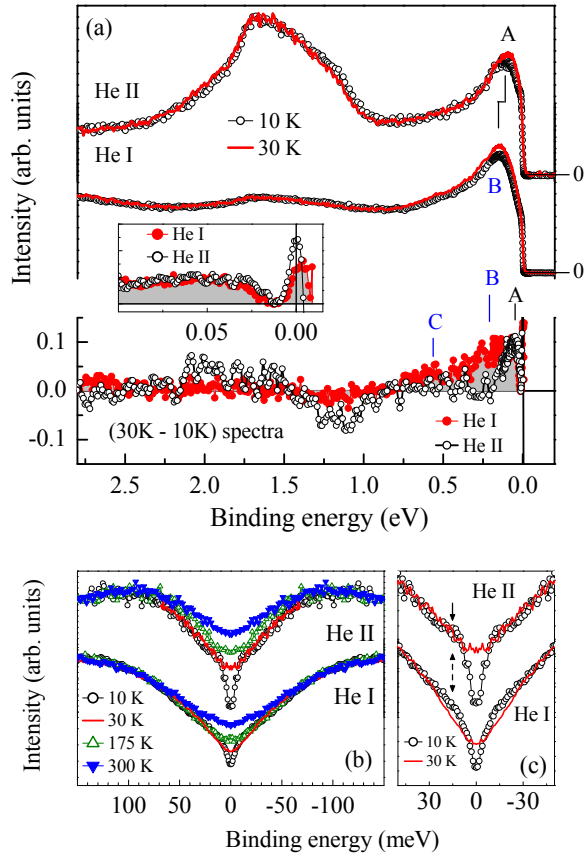}
 \vspace{-2ex}
\caption{(color online) (a) High resolution He {\scriptsize I} and
He {\scriptsize II} spectra at 10 K (symbols) and 30 K (lines). The
subtracted spectra (30 K - 10 K) at He {\scriptsize I} (solid
circles) and He {\scriptsize II} (open symbols) are shown in the
lower panel. The inset show the subtracted spectra near Fermi level.
(b) Symmetrized spectra at He {\scriptsize I} and He {\scriptsize
II} energies as a function of temperature. (c) Symmetrized 10 K and
30 K spectra near Fermi level.}
\end{figure}

We probed the evolution of the electronic structure of
EuFe$_2$As$_2$ as a function of temperature and photon energy
employing high energy resolution in the angle integrated
photoemission (AIPES) mode. The valence band using He {\scriptsize
I} ($h\nu$ = 21.2 eV) and He {\scriptsize II} ($h\nu$ = 40.8 eV)
photon energies at 10 K and 30 K are shown in Fig. 1(a). He
{\scriptsize II} spectra exhibit three distinct peaks around 0.1,
1.2 and 1.7 eV binding energies. The intensities of the 1.2 and 1.7
eV features become significantly weak in the He {\scriptsize I}
spectra relative to the 0.1 eV feature intensity. In addition, the
peak A in He {\scriptsize II} spectra is slightly shifted (feature
B) in the He {\scriptsize I} spectra. A change in photon energy from
He {\scriptsize I} to He {\scriptsize II} excitations leads to a
significant increase in photoemission cross-section of Eu 4$f$
states ($\sigma(Eu 4f)$) compared to all other contributions in the
valence band \cite{yeh-lindau}. Thus, the features in 1-2 eV energy
range can be attributed primarily to the Eu 4$f$ photoemission
\cite{JeevanPRB08} and the intensities near the Fermi level,
$\epsilon_F$ are due to Fe 3$d$ - As 4$p$ hybridized bands. The
change in lineshape and peak positions reflect the sensitivity of
the spectra to the photoemission cross section of constituent
electronic states.

Change in temperature across the AFM transition of the Eu moments
leads to a decrease in intensity near $\epsilon_F$ - this is evident
in the subtracted spectra (30 K - 10 K) shown in the lower panel of
Fig. 1(a) exhibiting three distinct features, A, B and C. The
feature, A appearing around 50 meV (see inset for clarity) has
similar intensity in both He {\scriptsize I} and He {\scriptsize II}
spectra suggesting mixed character of these states. The feature, B
at about 200 meV is strong in He {\scriptsize I} spectrum and absent
in He {\scriptsize II} spectrum. At He {\scriptsize I} and He
{\scriptsize II} energies, the values of $\sigma(Fe 3d)$ are 4.833
and 8.751, and that of $\sigma(As 4p)$ are 3.856 and 0.2949,
respectively\cite{yeh-lindau}. Clearly, As 4$p$ is significantly
stronger in He {\scriptsize I} energies, while Fe 3$d$ contributions
are stronger in He {\scriptsize II} energies. This indicates
dominance of As 4$p$ character in feature B. The feature, C
appearing around 0.5 eV is weak and observed in the He {\scriptsize
I} spectrum indicating its As 4$p$ character. The intensity changes
beyond 1 eV is visible primarily in He {\scriptsize II} spectrum
presenting the changes in Eu 4$f$ intensities. Clearly, the
electronic states near $\epsilon_F$ possess finite As 4$p$
contributions due to Eu 4$f$-As 4$p$ and Fe 3$d$-As 4$p$
hybridizations, and are highly sensitive to the AFM ordering.
Similar intensity of the feature A indicates participation of Fe
3$d$ states in the AFM ordering.

In Fig. 1(b) and 1(c), we investigate the spectral density of states
(SDOS) obtained by symmetrization ($I(\epsilon-\epsilon_F) +
I(\epsilon_F-\epsilon)$) of the experimental data, which provides a
good estimation of the intensity at $\epsilon_F$. The SDOS at both
the photon energies (He {\scriptsize I} and He {\scriptsize II})
exhibit significant reduction of intensity at $\epsilon_F$ across
SDW transition. The high energy resolution enabled to reveal an
additional gap at $\epsilon_F$ across the AFM transition at 20 K and
a quasiparticle peak appears around 20 meV in both He {\scriptsize
I} and He {\scriptsize II} spectra as evidenced in the rescaled
spectra in Fig. 1(c). While AFM transition can lead to a
gap/depletion at $\epsilon_F$ \cite{APLRSSingh,Bindu}, the spectral
evolution exhibiting a pseudogap-quasiparticle peak structure in
Fig. 1(c) resembles the widely known spectral evolution in various
unconventional superconductors exhibiting precursor effects
\cite{cuprates,APL-lab6} noting the fact that EuFe$_2$As$_2$ is
highly susceptible to show superconductivity as it exhibits
superconductivity under application of pressure and/or suitable
substitution at any of the three sites.

Various angle resolved photoemission (ARPES) studies of
EuFe$_2$As$_2$ showed that the temperature change across the SDW
transition ($\sim$ 190 K) leads to the formation of tiny hole and
electron pockets near ($\pi$,$\pi$) \cite{zhouPRB10}. The overlap of
the back-folded and non-folded bands in the SDW phase forms small
`droplets' in the Fermi surface \cite{Fink-EPL10}. Both, the bulk
measurements \cite{JeevanPRB08} and ARPES studies \cite{zhouPRB10}
suggested that Eu and FeAs sublattices are nearly decoupled as no
noticeable change was observed across the antiferromagnetic
ordering. However, the high energy resolution employed in this study
helped to reveal strong evidence of the influence of
antiferromagnetic transition on the conduction electrons responsible
for the electronic properties. In order to probe this in more
detail, we study the energy bands around the $\Gamma$ point along
(0,0)-($\pi$,$\pi$) direction as a function of temperature and
photon energy.

\begin{figure}
 \vspace{-4ex}
\includegraphics[scale=0.45]{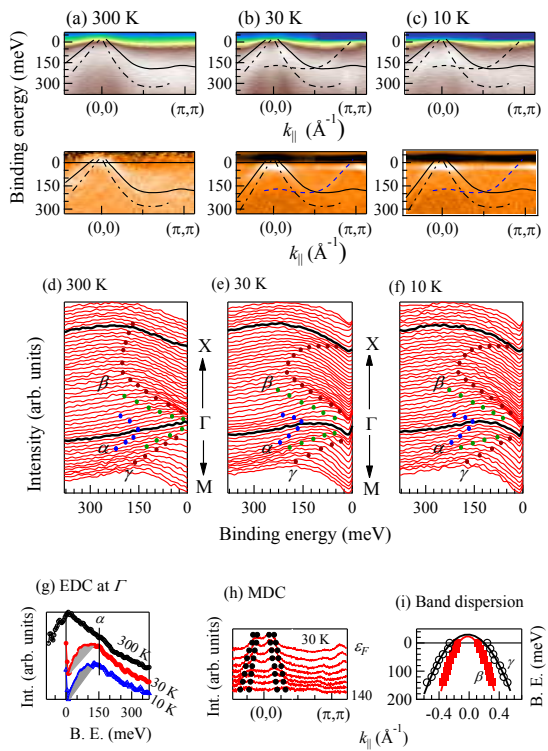}
 \vspace{-4ex}
\caption{(color online) He {\scriptsize II} ARPES data along
(0,0)-($\pi$,$\pi$)-direction at (a) 300 K, (b) 30 K and (c) 10 K.
The lower panel show the second derivative of the same data. Lines
are guide to the eye. The corresponding EDCs at (d) 300 K, (e) 30 K
and (f) 10 K. (g) EDC at $\Gamma$ at different temperatures. (h) MDC
at 30 K. The points in (h) and (i) represent the peak positions
obtained by fitting MDCs shown in (h).}
\end{figure}

In Fig. 2(a), 2(b) and 2(c), we show the He {\scriptsize II} ARPES
data at 300 K, 30 K and 10 K, respectively - the lower panels show
their second derivative. The band dispersions at 300 K (paramagnetic
phase) are consistent with the band structure results
\cite{zhouPRB10}. At 30 K, the region around ($\pi$,$\pi$) modifies
significantly - the dispersion around ($\pi$,$\pi$) becomes very
similar to that at $\Gamma$ point indicating band folding due to the
supercell formation in the SDW phase evidencing signature of a hole
pocket around ($\pi$,$\pi$). We calculated SDOS by dividing the
experimental spectra by the resolution broadened Fermi-Dirac
distribution function. The corresponding energy distribution curves
(EDCs) are shown in Fig. 2(d), 2(e) and 2(f) for the spectra at 300
K, 30 K and 10 K, respectively. Three distinct bands denoted by
$\alpha$, $\beta$ and $\gamma$ are discernible in the spectra. While
the $\alpha$ band folds back around 150 meV binding energy at
$\Gamma$ point, $\beta$ and $\gamma$ bands cross $\epsilon_F$
forming hole pockets around $\Gamma$ point
\cite{zhouPRB10,Fink-EPL10}.

The energy band dispersions of $\beta$ and $\gamma$ bands were
derived via fitting the momentum distribution curves (MDC). The peak
positions are shown in Fig. 2(h) and 2(i). The estimated values of
$k_F$ for the $\beta$ and $\gamma$ bands are about 0.1 \AA$^{-1}$
and 0.2 \AA$^{-1}$, respectively. The effective mass ($m^\star/m_e$
= $\hbar^2/(\partial^2\epsilon/\partial k^2)$) of the charge
carriers is found to be 2.5 and 6.1 for the $\beta$ and $\gamma$
bands, respectively, consistent with the values estimated from
quantum oscillation studies in similar systems \cite{QuOsc}. The
electronic states forming the $\gamma$ band are found to be more
correlated than others and play active role in the SDW transition.
The Fermi velocities for the two bands are 6.5$\times$10$^4$ and
4.5$\times$10$^4$ m/sec, respectively. These values remain almost
the same across the antiferromagnetic transition at 20 K.

\begin{figure}
 \vspace{-4ex}
\includegraphics[scale=0.45]{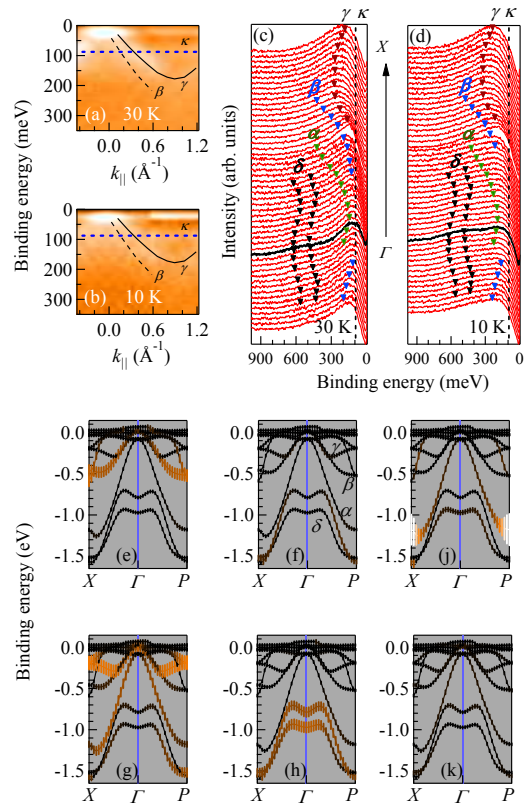}
 \vspace{-4ex}
\caption{(color online). Second derivative of the He {\scriptsize I}
ARPES data along $\Gamma$M-direction at (a) 30 K and (b) 10 K. EDCs
of the SDOS of corresponding spectra at (c) 30 K and (d) 10 K.
Calculated energy bands near Fermi level highlighting the (e) Fe
3$d_{xy}$, (f) Fe 3$d_{x^2-y^2}$, (g) Fe 3$d_{xz}$+3$d_{yz}$, (h) Fe
3$d_{z^2}$, (j) As 4$p_z$ and (k) As 4$p_x$+4$p_y$. The intensity of
the As 4$p$ states are magnified by 10 times.}
\end{figure}

The EDC around $\Gamma$ point exhibit interesting evolution with
temperature. At 300 K, EDC exhibits large intensity near
$\epsilon_F$. The 30 K data, on the other hand, exhibits a dip at
$\epsilon_F$, which becomes more prominent at 10 K consistent with
the AIPES data. This is shown with better clarity in Fig. 2(g),
where we show the EDCs at $\Gamma$ point at various temperatures.
The $\alpha$ band appearing around 150 meV remains almost unchanged
in the whole temperature range studied. An additional feature (see
the shaded area) appears around 80 meV in the 30 K spectrum that
reduced significantly in the 10 K data. To get a better
understanding of this feature, we investigated the He {\scriptsize
I} spectra along the same $k$-direction at 30 K and 10 K in Fig. 3.
The $\beta$ and $\gamma$ bands remain two dimensional in the whole
temperature range and hence, the difference in the values of $k_z$
probed by He {\scriptsize I} ($\sim$10$\pi/c$) and He {\scriptsize
II} ($\sim$13$\pi/c$), will have negligible influence on the
dispersions \cite{kaminskii-2Dto3D}. The dispersion of the $\beta$
and $\gamma$ bands are evident in Fig. 3(a) and 3(b), and in the
EDCs of the SDOS shown in Fig. 3(c) and 3(d) as well. In addition, a
weakly dispersing feature around 80 meV denoted by $\kappa$ is
observed.

We have calculated the energy bands and their orbital characters
using {\it state-of-the-art} full potential linearized augmented
plane wave method within the local density approximations using
Wien2k software \cite{wien2k}. The calculated energy bands
corresponding to the experimental data are shown in the figure,
where the thickness of the lines represent the orbital character to
be 3(e) Fe 3$d_{xy}$, 3(f) Fe 3$d_{x^2-y^2}$, 3(g) Fe
3$d_{xz}$+3$d_{yz}$, 3(h) Fe 3$d_{z^2}$, 3(j) As 4$p_z$ and (k) As
4$p_x$+4$p_y$ derived. The calculated results, consistent with the
results in literature for various similar systems, resemble well the
experimental data. The $\delta$ bands possessing $d_{z^2}$ character
appear at higher binding energies. The $\alpha$ and $\beta$ bands
possess ($d_{xz},d_{yz}$) and $d_{xy}$ character, respectively and
are less influenced by the magnetic transitions - a small shift
($\sim$ 20 meV) of the $\beta$ band is observed across the AFM
transition at 20 K. Fig. 3(g) indicates ($d_{xz},d_{yz}$) symmetry
of the $\gamma$ band, which is responsible for the SDW transition.
This is understandable as the As layers appear above and below the
Fe-layers and hence, the hybridization mediating inter-cite coupling
will be dominated by ($d_{xz},d_{yz}$) electronic states.

\begin{figure}
\includegraphics[scale=.45]{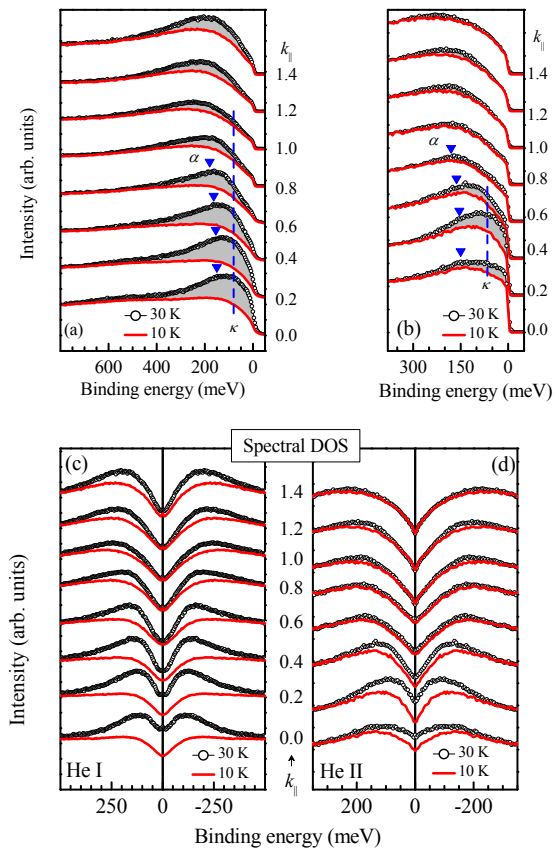}
\caption{(color online). EDCs at different $k$-points in the (a) He
{\scriptsize I} and (b) He {\scriptsize II} spectra. The shaded area
show the change across antiferromagnetic transition. The SDOS at 10
K (lines) and 30 K (open circles) obtained by symmetrization of the
raw data at different k-points at (c) He {\scriptsize I} and (d) He
{\scriptsize II} photon energies.}
\end{figure}

Although the energy bands in the vicinity of $\epsilon_F$ possess
dominant Fe 3$d$ character, the contribution of As 4$p$ states is
significant due to hybridization among the corresponding electronic
states. In order to investigate the feature, $\kappa$ further, we
compare the EDCs at 30 K and 10 K observed in He {\scriptsize I} and
He {\scriptsize II} spectra in Fig. 4. While the photoemission
intensities reduce with temperature in both the photon energies, the
largest change is observed in the He {\scriptsize I} spectra. The
relative photoemission cross section of the As 4$p$ states is
enhanced significantly at He {\scriptsize I} energy compared to that
at He {\scriptsize II} energy \cite{yeh-lindau}. Thus, the feature
$\kappa$ can be attributed primarily to As 4$p$ states. Weak
dispersion suggests hybridization of these states with the weakly
dispersing Eu 4$f$ states. This characterization is further
justified by the fact that the intensity of the $\kappa$ band
reduces significantly with the decrease in temperature across the
AFM transition temperature due to Eu 4$f$ moments. The symmetrized
Spectral DOS are shown in Figs. 4(c) and 4(d) for He {\scriptsize I}
and He {\scriptsize II} excitations, respectively. The largest
change in SDOS is observed at $k$-points, where the $\beta$ and
$\gamma$ bands cross the Fermi level - the intensities at all other
$k$-points are almost identical. The change in intensity is largest
in the He {\scriptsize I} spectra indicating again large As 4$p$
character of these electronic states.

In summary, we studied the temperature evolution of the electronic
structure of EuFe$_2$As$_2$, an ideal compound to study the
interplay between magnetism and superconductivity generally known to
be two mutually exclusive phenomena. The signature of
pseudogap-quasiparticle feature, typical of various superconducting
materials is manifested in the high resolution data at low
temperatures; these features possess large As 4$p$ character. The
angle resolved data demonstrate band folding due to the spin density
wave transition. Distinct signature of an additional weakly
dispersing feature is manifested in the low temperature data
possessing large As 4$p$ character that reduces in intensity across
antiferromagnetic transition in the Eu layer. Evidently, the Eu
4$f$-As 4$p$ and Fe 3$d$-As 4$p$ hybridizations are significant and
play major role in deriving the electronic properties. These results
demonstrate evidence of a link of antiferromagnetism and pseudogap
states that could be revealed due to the employment of high
resolutions.

The authors KM and NS acknowledge the financial support from the
Department of Science and Technology, Government of India under the
`Swarnajayanti Fellowship programme'.

\end{document}